\documentclass[aps,amssymb,prl,twocolumn,showpacs]{revtex4}
\usepackage{graphics}

\begin{document}
\bibliographystyle{prsty}

\title{Subextensive Scaling in the Athermal, Quasistatic Limit of Amorphous Matter in Plastic Shear Flow}
\author{Craig Maloney$^{(1,2)}$}
\author{Ana\"el Lema\^{\i}tre$^{(1,3)}$}
\affiliation{
$^{(1)}$ Department of physics, University of California, Santa Barbara, California 93106, U.S.A.}
\affiliation{$^{(2)}$ Lawrence Livermore National Lab - CMS/MSTD, Livermore, California 94550, U.S.A.}
\affiliation{$^{(3)}$ L.M.D.H. - Universite Paris VI, UMR 7603,
4 place Jussieu, 75005  Paris - France}
\date{\today}

\begin{abstract}
We present the results of numerical simulations of an atomistic system undergoing plastic shear flow in the athermal, quasistatic limit.  The system is shown to undergo cascades of local rearrangements, associated with quadrupolar energy fluctuations, which induce system-spanning events organized into lines of slip oriented along the Bravais axes of the simulation cell.  A finite size scaling analysis reveals subextensive scaling of the energy drops and participation numbers, linear in the length of the simulation cell, in good agreement with the observed real-space structure of the plastic events.
\end{abstract}
\maketitle

The recent years have seen an important number of numerical and theoretical
studies of plasticity in amorphous materials. 
The microscopic picture of plastic deformations which emerges from these
studies, however, is still incomplete... at best, fragmented.
Numerical evidence that plastic deformation involves 
highly heterogenous displacements of molecules led, early on, to the concept of
``shear transformation zones'', which are expected to play,
for amorphous solids, the r\^ole of defects in crystals .~\cite{MaedaT78,Argon79,FalkL98}
Most theoretical models of plasticity rely on this idea and,
following Eshelby,~\cite{Eshelby57} on the expectation that elementary shear
transformations are associated with quadrupolar energy fluctuations.
Theoretical works indicate that the existence of quadrupolar elastic fields,
and the consequent long-range interactions between shear transformation zones 
is an important mechanism that can induce strain localization and fracture 
in amorphous materials.~\cite{BulatovA94a,Langer01,BaretVR02}
Although quadrupolar energy fluctuations have been observed in 
a numerical model of dry foams,~\cite{KablaD03} they have never been seen 
in molecular systems.

\begin{figure}
\rotatebox{-90}{\resizebox{!}{.45\textwidth}{{\includegraphics{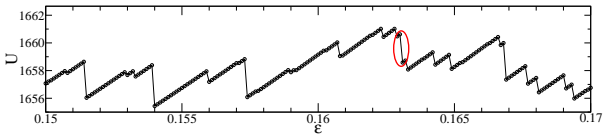}}}}
\caption{Potential energy as a function of strain during quasi-static shear of a 200x200 system.} \label{fig:trajectory}
\end{figure}

This line of research should be contrasted with the phase space interpretation 
of plastic deformation recently proposed by Malandro and Lacks,~\cite{MalandroL99} 
on the basis of the inherent structure formalism.~\cite{goldstein69}
These authors study shear induced changes in the potential energy landscape, 
and the consequences of such changes on the macroscopic mechanical behavior of glasses.
In order to isolate these effects, Malandro and Lacks consider the quasi-static 
deformation of an amorphous material at zero-temperature, 
a protocol which has been used since early numerical studies 
as a means to bypass intrinsic 
limitations of molecular dynamics algorithms.~\cite{MaedaT78}
For small deformations, the system follows shear induced changes
of a local minimum (inherent structure) in the potential energy landscape.
Elementary catastrophic events occur when the local minimum in which the system 
resides annihilates during a shear-induced collision with a saddle point.
The deformation of an amorphous material thus involves a series 
of reversible (elastic) branches intersected by plastic rearrangements 
(see figure~\ref{fig:trajectory}).

The inherent structure formalism provides a precise definition 
of an elementary plastic rearrangement,
but several questions arise about the spatial organization of 
these transitions:
Are plastic events related to shear transformation zones
and quadrupolar energy fluctuations?
Do they involve spatially localized dynamical structures?
If not, how do these structures scale with system size?
Conflicting answers to these questions can be found in the literature.
From measurements of participation ratio, Malandro and Lacks indicate that 
the elementary rearrangements they observe are localized.~\cite{MalandroL99}
Durian and coworkers, for a model of foam (athermal by construction), 
observe a power-law distribution of energy drops at small strain rates, 
but with a system-size independent cut-off,
indicating that no scaling behavior is to be seen, unless at a very specific
point in the jamming phase-diagram.~\cite{TewariSDKLL99}
A contradictory viewpoint is supported by the observations by Yamamoto and Onuki
of increasing lengthscale suggesting the emergence of delocalized events, 
and critical behavior in the low-temperature, low strain-rate limit.~\cite{YamamotoO98}.

In this work, we study spatial organization of elementary 
transitions between inherent structures in quasi-static shear deformation, 
using the soft-sphere interaction potential used by Durian.~\cite{Durian97}
This choice was initially motivated by our intent to avoid
delocalized structures. With this model we observe:
(i) quadrupolar energy fluctuations and cascades of these
during single transitions between inherent structures,
(ii) elongated, crack-like events that span 
the whole shear cell.
Next (iii) we measure the distribution of energy drops and participation 
ratios and show that the size of typical events scales linearly with the 
length of the simulation cell.
We thus show that scaling behavior is to be found in the quasi-static
limit, and that it is associated with a cascade of spatially correlated 
quadrupolar energy fluctuations, reminiscent of theoretical 
considerations.~\cite{BaretVR02}


We perform numerical experiments using simple shear, or so called Lees-Edwards, boundaries.  Particles interact through the soft sphere-potential~\cite{Durian97}:
\[U_{ij}=
\left\{
\begin{array}{ll}
  \frac{1}{2}k \left(1-d_{ij}\right)^{2} & d_{ij}\leq 1 \\
  0 & d_{ij}>1
\end{array}
\right\}
\]
where,
$d_{ij}=2\left(\|\vec{x_{i}}-\vec{x_{j}}\|\right)/\left(D_{i}+D_{j}\right)$, with $D_{i}$, the diameter of particle $i$.
$k$, being the only energy scale in the problem is set to unity.  In order to prevent crystallization, a binary mixture is used with: $D_{L}=1, D_{S}=D_{L}\frac{\sin{\frac{\pi}{10}}}{\sin{\frac{\pi}{5}}}, N_{L}/N_{S}=\frac{1+\sqrt{5}}{4}$~\cite{FalkL98} where $N_{L}(N_{S})$ is the number of large (small) particles.
The athermal, quasistatic shear algorithm consists of two parts.  First the simulation cell is deformed by a small amount with particle positions fixed in reciprocal space (i.e. fixed relative to the Bravais axes of the simulation cell), producing an affine deformation in real space.  Next the potential energy of the system is minimized with the shape of the simulation cell held fixed resulting in corrections to the affine deformation.  
Physically, the quasistatic algorithm corresponds to a material which is being sheared in a much shorter time than the thermally induced structural relaxation time, but a much longer time than any microscopic times: $\tau_{\text{micro.}}\!<<\!\frac{1}{\dot{\gamma}}\!<<\!\tau_{\text{struc.}}$.

The initial sample is prepared with a standard conjugate gradient minimization applied to a random state. 
A fixed area simulation cell with a packing fraction of $1.0$ is used for all systems.  This density, well above the random close packing limit, was originally thought to preclude the emergence of non-localized structures.~\cite{TewariSDKLL99} We use a strain step of size $10^{-4}$ for all simulations.  Results will be presented for three ensembles of systems (with sizes: $L^2=$ 12.5x12.5, 25x25, and 50x50) and also one single 200x200 system.

Proceeding with a discussion of the single 200x200 system, figure~\ref{fig:trajectory} shows the potential energy of the relaxed configurations as a function of the applied shear strain for a small interval of strain from $.15$ to $.17$.  The curve is composed of continuous segments, broken up by discontinuous drops.  Malandro and Lacks~\cite{MalandroL99} have demonstrated that each discontinuity arises from the destruction of a potential energy minimum induced by the imposed shear strain.  In agreement with them, we find that the system is microscopically reversible upon changing the sense of the strain during the continuous segments, but becomes irreversible across the discontinuities which constitute the fundamental plastic events.

\begin{figure}
\rotatebox{-90}{\resizebox{!}{.45\textwidth}{{\includegraphics{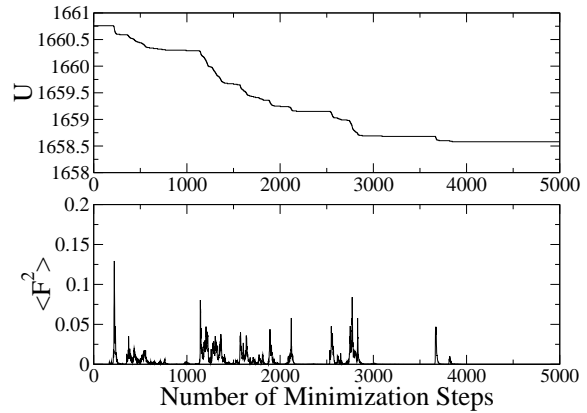}}}}
\caption{Potential energy and sum of the squares of the forces as the system progresses through the minimization algorithm during the event circled in figure~\ref{fig:trajectory}, above.} \label{fig:descent}
\end{figure}

Next we look at the energy relaxation during a typical plastic event, which is circled in figure~\ref{fig:trajectory}. 
The energy and sum of the squares of the forces during the conjugate gradient descent for this single energy minimization are shown in figure~\ref{fig:descent}.  In these plots, the horizontal axis represents the amount of progress through the conjugate gradient algorithm.  We checked in smaller systems that minimization via integrating the equations of steepest descent yields similar curves.  In this latter case the horizontal axis can be directly interpreted as time, but steepest descent cannot be used in large systems due to its intrinsic inefficiency. In steepest descent dynamics, the time derivative of the energy is precisely the sum of the squares of the forces, and this relation holds reasonably well for our conjugate gradient trajectories. Figure~\ref{fig:descent} shows plateaus in the energy, which correspond to configurations where the forces are small and hence are very close to being mechanical equilibria. These configurations are {\it quasi-equilibria\/}: deficient equilibria, each of which allows for an escape into a new quasi-equilibrium with lower energy.  The system cascades through a series of quasi-equilibria of decreasing energy until finally arresting in a true minimum.

\begin{figure}
\begin{minipage}{.2\textwidth}
\rotatebox{0}{\resizebox{!}{\textwidth}{{\includegraphics{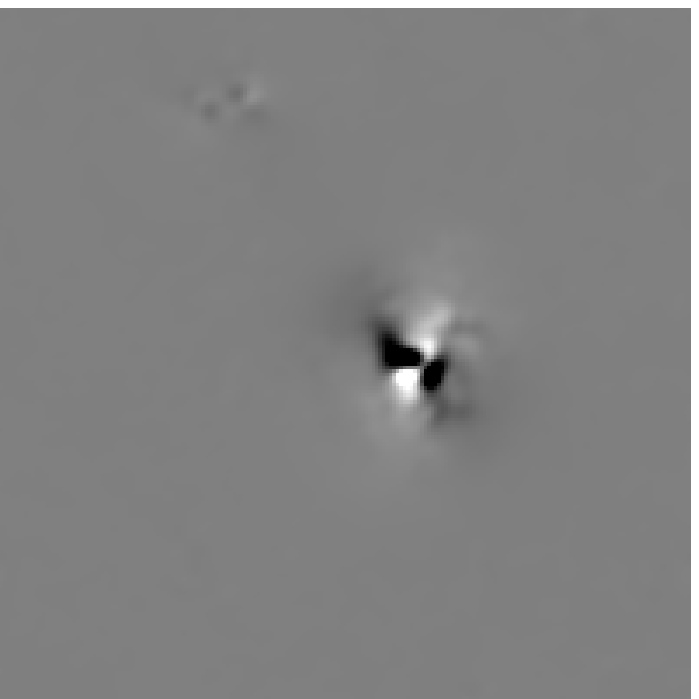}}}}
\end{minipage}
\begin{minipage}{.2\textwidth}
\rotatebox{0}{\resizebox{!}{\textwidth}{{\includegraphics{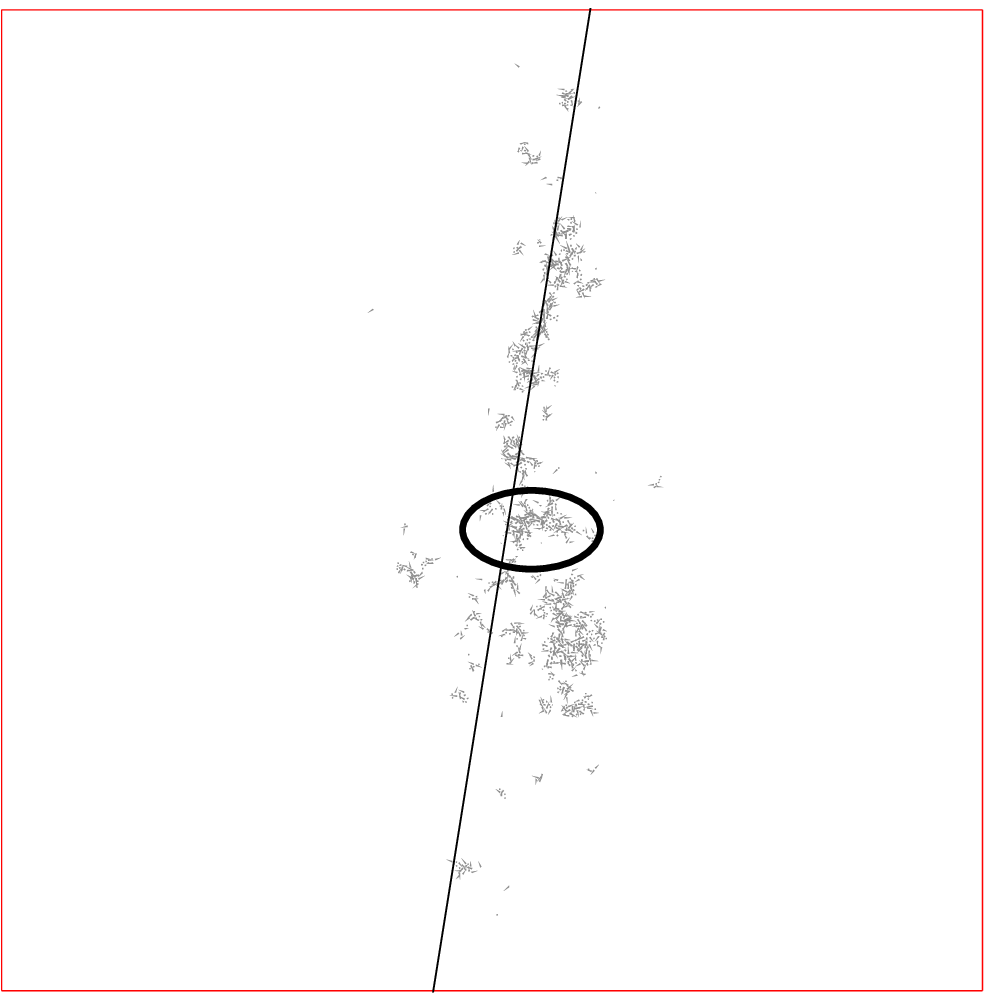}}}}
\end{minipage}
\caption{a) The change in potential energy during the first force peak in figure~\ref{fig:descent}.  The color scale is linear with pure white (black) corresponding to a local decrease (increase) in energy per unit area equal to: $5\times 10^{-4}$.  The orientation of this quadrupole is precisely what one would expect based on the direction of the principle axes of the applied shear strain. b) The local relative displacements (defined as the displacement of a particle with respect to the average displacement of its neighbors) that result from the entire cascade circled in figure~\ref{fig:trajectory}.  Only displacement vectors which are larger than $.1$ times the maximum are drawn.  The black line is a guide to the eye oriented along the oblique Bravais axis.  The particles inside the circled cluster are those which move during the first peak in the forces from figure~\ref{fig:descent} and produce the quadrupolar pattern shown in a). \label{fig:quadrupole}}
\end{figure}

The observation of these transitions suggests that a typical plastic event, as circled on figure~\ref{fig:trajectory}, might be decomposed into {\em elementary sub-events}.  Figure~\ref{fig:quadrupole}a shows the resulting change in potential energy which occurs during the first force peak (in figure~\ref{fig:descent}).  
The quadrupolar pattern is apparent.
It is the first time such a field has been observed in an atomistic simulation.  We emphasize that the energy dissipation field shown in figure~\ref{fig:quadrupole}a corresponds to a \emph{single} elementary sub-event and contrast this with the work of Kabla and Debr\'egeas~\cite{KablaD03} who observe such a quadrupolar field in a mechanical film model only after averaging over many plastic events.  We observe these quadrupoles \emph{generically} during the onset of other typical events like the one circled in figure~\ref{fig:trajectory}, however, after the onset the situation becomes more complex as a cascade is initiated in which the system proceeds through a series of such elementary sub-events. The spatial signature of each elementary sub-event is, of course, noisy and, as the cascade proceeds, several elementary sub-events may occur concurrently and overlap in space. These effects contribute to an increasingly complicated energy dissipation field in which it is often difficult to disentangle elementary quadrupolar patterns. Our observations, however, are consistent with viewing every cascade as a superposition of quadrupolar fields, each associated with an elementary sub-event. 

Turning to the spatial organization of the sub-events into a cascade,  the local relative displacements of each particle are drawn in figure~\ref{fig:quadrupole}b, where the particles associated with the onset are circled.  In this picture, non-affine rearrangements cluster along the oblique Bravais axis of the simulation cell. We observe similar crack-like patterns generically in plastic events, aligned preferentially along the oblique or non-oblique (horizontal) Bravais axes. These patterns are reminiscent of the shear-bands which are predicted by several mesoscopic models of plasticity, and are expected to emerge from the interactions of local shear transformations mediated by quadrupolar fluctuations in the elastic field~\cite{Eshelby57,BulatovA94a,BaretVR02,Langer01}. We emphasize, however, that the patterns we observe here are certainly distinct from \emph{persistent} shear bands: they are transient events which occur during a \emph{single, infinitesimal} strain step. The location and orientation of these patterns fluctuate as the system is sheared, and by no means can we identify the emergence of any stationary heterogeneous deformation field, as observed in some molecular simulations~\cite{RottlerR03}.

The existence of non-localized dynamical structures is consistent 
with the observation by Yamamoto and Onuki,~\cite{YamamotoO98} 
in molecular dynamics simulations of glassforming systems,
of an increasing correlation length in the limit where 
first temperature and then strain rate go to zero. 
They claim that their data is consistent with the existence 
of critical behavior in this limit,
but were unable to access the putative critical point due to inherent
limitations in the molecular dynamics algorithm.
Our algorithm is locked precisely at the $T\to0, \dot\gamma\to0$ limit,
which enables us to perform finite-size scaling analysis \emph{at} this point.
Before proceeding, we must emphasize that we study here a different molecular 
model which was not expected to display non-localized structures.~\cite{Durian97} 
Moreover, we now show that the crack-like patterns we observe are responsible 
for the emergence of specific types of scaling
which were not identified in previous numerical works.


\begin{figure}
\rotatebox{-90}{\resizebox{!}{.4\textwidth}{{\includegraphics{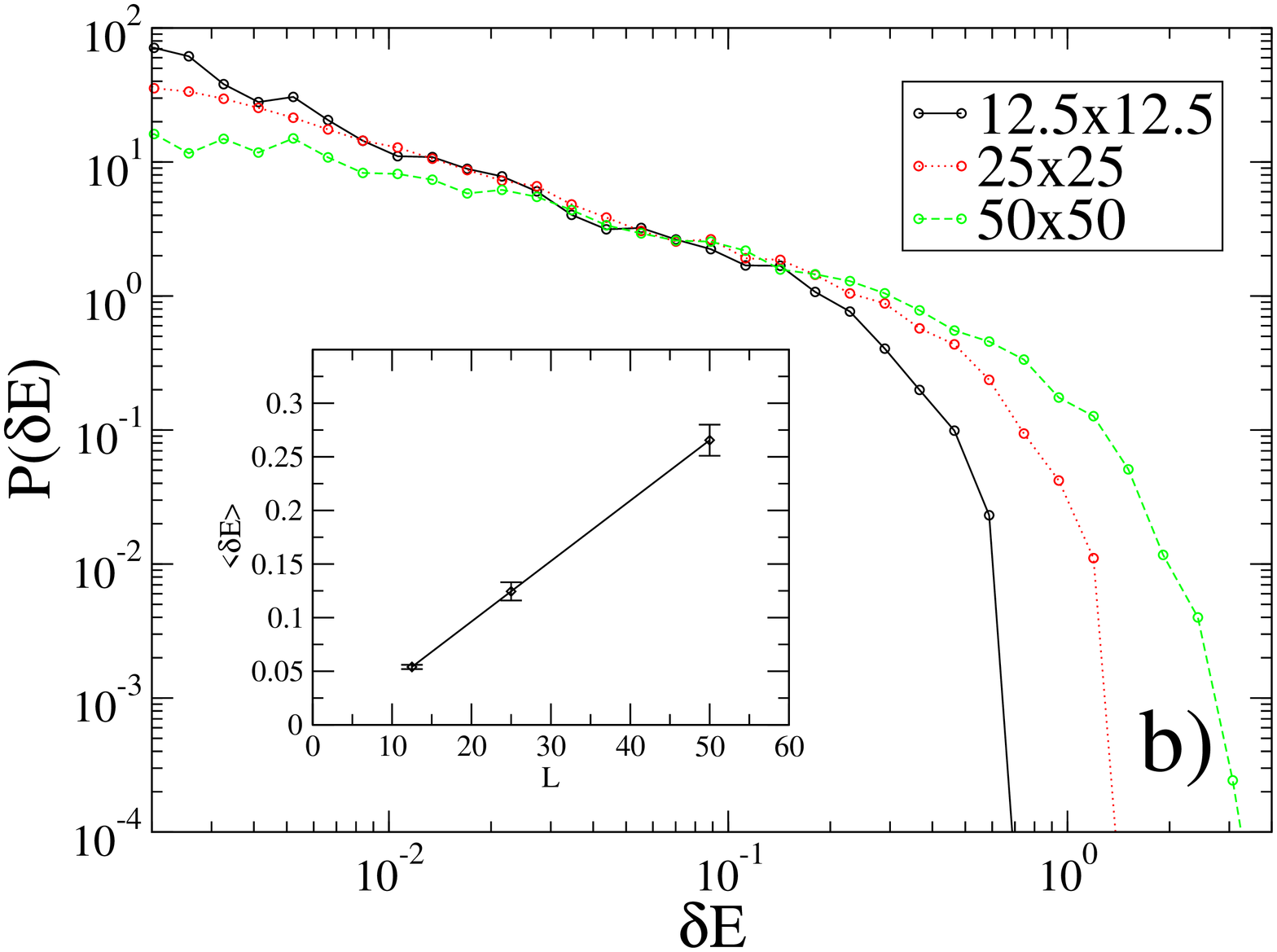}}}}
\rotatebox{-90}{\resizebox{!}{.4\textwidth}{{\includegraphics{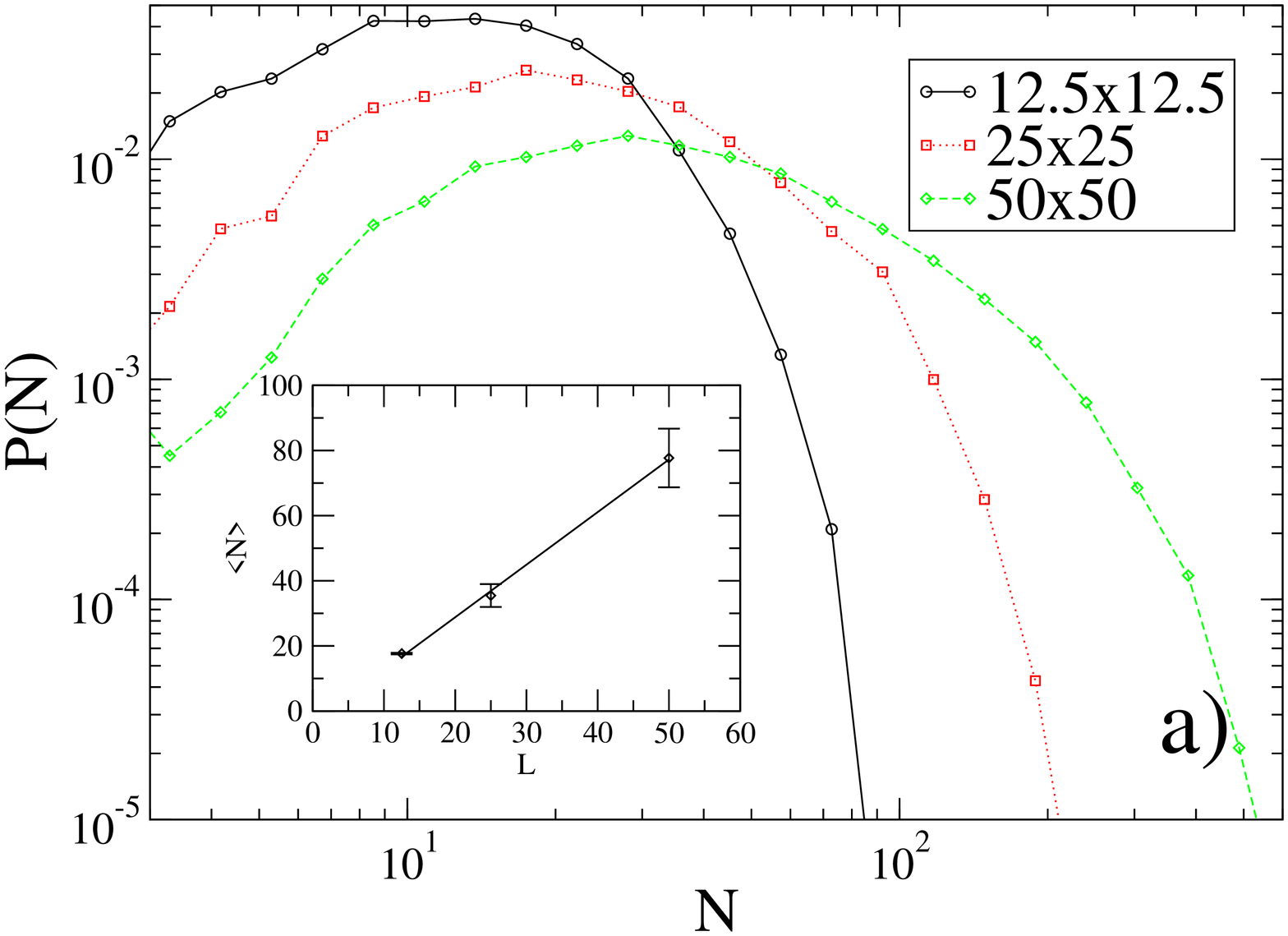}}}}
\caption{a),b) Distribution of participation number and energy drops for systems of lengths, $12.5$,$25$,and $50$. Insets: Scaling of the average participation number and energy drop with system length.} \label{fig:probDist}
\end{figure}

Results for the steady-state distribution of energy drops and 
participation number are given in figure~\ref{fig:probDist} for different system sizes.
We first examine the distribution of participation numbers $N$. We find a clear increase of the average with the linear system size, $\langle N\rangle\sim L$, and a corresponding shift in the distribution. This scaling is consistent with the assumption that the dominant events are system spanning faults of length $L$ with a typical transverse length scale $a$ which does not depend on system size. We next look at the distribution of energy drops: it is well described by a power law with an exponential cutoff at large events.  The power seems to be slightly smaller for our largest system, and ranges from $.7$ to $.5$, which is in rough accord with earlier results~\cite{Durian97,TewariSDKLL99}. However, as was the case with the participation number, we observe dramatic system size effects, with the average energy and the cut-off increasing linearly with $L$. This scaling is consistent with the idea that the energy dissipated during a plastic event scales like $\langle E\rangle\sim E_0 \langle N \rangle\sim E_0 L$, where $E_0$ is the elementary energy released per quadrupolar fluctuation.


Our observations differ from previous claims found in the literature.
The distribution of participation numbers was studied by Malandro and Lacks for 
a three-dimensional model of a glass.~\cite{MalandroL99}
These authors concluded that the average participation number became 
independent of system size for large systems, but their data only disfavors 
an extensive ($L^{3}$) scaling. From our observations, we expect that, 
in three dimensions, plastic events are likely to organize 
into fault planes: in such a case, the scaling would become
$\langle N\rangle\sim L^{2}$, which is also consistent with 
Malandro and Lacks' data.~\cite{MalandroL99}
Here we come to an important point: the non-extensivity of the participation 
number does not mean that structures are localized, 
as sub-extensive, system-spanning, structures may emerge.

Our conclusion also differs from that of Tewari {\it et. al. }~\cite{TewariSDKLL99}, 
on two-dimensional models of foams. We believe this discrepancy originates from 
a subtle consequence of their use of finite strain rate simulations. 
In quasistatic shear, 
the respective timescales of plastic events and shear are completely separated.  
However, the crack-like patterns we observe result from {\it cascading} sub-events,
as information propagates through the system;
the plateaus in figure~\ref{fig:descent} correspond 
to times when information propagates with little dissipation.
At finite strain rate, the spatial development of plastic events and 
the overall deformation of the material occur concurrently:
the plateaus of figure~\ref{fig:descent}, become slightly tilted due to the overall 
energy increase induced by the finite strain-rate.
The criterion Tewari {\it et. al.} use to separate individual plastic events
stipulates that energy should decrease monotonically during a single event;
it may misidentify quasi-equilibria for true equilibria, 
thus precluding the complete identification of elementary events.
A finite shear rate may thus artificially ``break'' single plastic rearrangements into 
several spurious sub-events which have no simple interpretation in the 
energy landscape.


In conclusion, we have presented results on an atomistic system sheared in the athermal quasistatic limit.  
We demonstrated the organization, during cascades, of elementary quadrupolar plastic zones into lines of slip oriented along the Bravais axes of the cell.  
We proceeded to perform a finite size scaling analysis which revealed a linear system size dependence which indicates that the faultlike patterns of energy fluctuations play a major r\^ole in the emergence of scaling behavior. 
%
The overall picture which emerges from our simulations thus explains and clarifies various controversial claims found in the literature, providing a first step toward a unified view of amorphous plasticity based on both the energy landscape and real space pictures.

This work was partially supported under the auspices of the U.S. Department of Energy by the University of California, Lawrence Livermore National Laboratory under Contract No. W-7405-Eng-48, by the NSF under grants DMR00-80034 and 
DMR-9813752, by the W. M. Keck Foundation, and EPRI/DoD through the Program on Interactive Complex Networks.  CM would like to acknowledge the guidance and support of V.~V. Bulatov and J.~S. Langer and the hospitality of the LLNL University Relations program and HPCMS group.


\end{document}